\documentclass[letter]{aa}  
\usepackage{xcolor}
\usepackage{graphicx}
\usepackage[normalem]{ulem}
\usepackage{txfonts}

\usepackage{hyperref}
\hypersetup{
    linkcolor=blue,
    filecolor=magenta,      
    urlcolor=cyan,
    }

\begin{document} 
   \title{Time-dependent response of protoplanetary disk temperature to an FU Ori-type luminosity outburst}

   \author{S.I. Laznevoi
          \inst{1},
          V.V. Akimkin\inst{2},
          Ya.N. Pavlyuchenkov\inst{2},
          V.B. Il'in \inst{1,3},
          \'A. K\'osp\'al \inst{4,5,6}
          and P. \'Abrah\'am \inst{4,5,7}
          }
\institute{Saint Petersburg State University, Universitetskij pr. 28, St. Petersburg 198504, Russia
\and Institute of Astronomy, Russian Academy of Sciences, Pyatnitskaya str. 48, Moscow 119017, Russia\\
\email{akimkin@inasan.ru}
\and Main Astronomical Observatory at Pulkovo, Pulkovskoe sh. 65/1, St. Petersburg 196140, Russia
\and Konkoly Observatory, HUN-REN Research Centre for Astronomy and Earth Sciences, MTA Centre of Excellence, Konkoly-Thege Mikl\'os \'ut 15-17, 1121 Budapest, Hungary
\and Institute of Physics and Astronomy, ELTE E\"otv\"os Lor\'and University, P\'azm\'any P\'eter s\'et\'any 1/A, 1117 Budapest, Hungary
\and Max-Planck-Insitut f\"ur Astronomie, K\"onigstuhl 17, 69117 Heidelberg, Germany
\and Department of Astrophysics, T\"urkenschanzstra{\ss}e 17, 1180, Vienna, Austria}
 
\titlerunning{Protoplanetary Disk Temperature Response to FU Ori-type Outburst}
\authorrunning{Laznevoi et al. }
\date{Received \today; accepted \today}

 
\abstract
   {The most prominent cases of young star variability are accretion outbursts in FU Ori-type systems. The high power of such outbursts causes dramatic changes in the physical and chemical structure of a surrounding protoplanetary disk. As characteristic thermal timescales in the disk are comparable to the duration of the outburst, the response of its thermal structure is inherently time dependent.}
{We  analyzed  how the disk thermal structure evolves under the substantial---yet transient---heating of the outburst. To cover different possible physical mechanisms driving the outburst, we examined two scenarios: one in which the increased accretion rate is confined to a compact sub-au inner region and the other  where it affects the entire disk.}
   {To model the disk temperature response to the outburst we performed time-dependent radiation transfer using the HURAKAN code. The disk structure and the luminosity profile roughly correspond to those of  the FU Ori system itself, which went into outburst about 90 years ago and reached a luminosity of 450 $L_\odot$. The static RADMC-3D code was used to model synthetic spectral energy distributions (SEDs) of the disk based on the temperatures calculated with HURAKAN.}
   {We find that optically thick disk regions require several years to become fully heated during the outburst and a decade to cool after it. The upper layers and outer parts of the disk, which are optically thin to thermal radiation, are heated and cooled almost instantaneously. This creates an unusual radial temperature profile during the early heating phase with  minima at several au both for the fully active and compact active disk scenarios. At the cooling phase, an unusual temperature gradient occurs in the vertical direction with the upper layers being colder than the midplane for both scenarios. Near- and mid-infrared SEDs demonstrate a significant and almost instantaneous rise by $1-2$ orders of magnitude during the outburst, while the millimeter flux shows a change of only a factor of a few, and is slightly delayed with respect to the central region luminosity profile. 
   }
   {}

   \keywords{protoplanetary disks -- accretion, accretion disks -- radiative transfer -- stars: pre-main sequence -- stars: variables: T~Tauri, Herbig Ae/Be}

   \maketitle

\section{Introduction}

FU Orionis-type luminosity outbursts are associated with a sudden increase in protostellar disk accretion \citep{1996ARA&A..34..207H, 2023ASPC..534..355F}. Although the outburst-triggering mechanism remains unclear \citep{2021A&A...647A..44V, 2024MNRAS.528.2182N}, the observed phenomena are thought to originate primarily in the inner active accretion disk, which is confined to within a few tenths of an au. Models of the active disk, incorporating complex physical processes, are rapidly advancing \citep[e.g.,][]{Zhu_2009, 2020MNRAS.495.3494Z, 2021A&A...655A.110S, 2024A&A...692A.171C}. Recent observations from  far-UV to centimeter wavelengths provide new valuable constraints on these models \citep{2017A&A...602A..19L, 2019ApJ...884...97L, 2022A&A...663A..86L, 2024ApJ...973L..40C}. The physics of outer disk regions are also studied in different aspects \citep[e.g.,][]{2024MNRAS.527.9655A, 2024A&A...687A.136C, 2024ApJ...966...96H, 2024A&A...686A.309Z}.

Luminosity outbursts of FU Ori-type systems should have a profound impact on the physical and chemical structure of their protoplanetary disks. For a typical outburst duration and magnitude ($\sim 10$ yr and $\sim100\,L_{\sun}$, respectively; \citealt{2023ASPC..534..355F}), the energy released in the inner active disk may significantly exceed the total disk thermal energy at the pre-outburst stage. In addition to significant restructuring of the inner active disk, its strong UV/optical radiation should cause profound changes in the thermal structure of the outer passive disk. The heating process is intrinsically time dependent: different parts of the passive disk are heated on different timescales as radiation diffuses toward the optically thick midplane. Similarly, when the central source luminosity drops back to normal, the disk cools nonhomogeneously with optically thick parts shedding   excess energy on timescales comparable with those of observational monitoring programs. 

The time-dependent nature of disk heating and cooling is currently poorly understood, and many popular radiation transfer codes for protoplanetary disks treat the process statically, i.e. assuming instantaneous adaptation of the disk temperature to the central region luminosity. Among the notable exceptions is the recent study by \citet{2024A&A...688A...8W}, where a time-dependent Monte Carlo radiation transfer code was used to model an accretion burst in a massive young stellar object. Another example is the work by \cite{2018A&A...617L...7S}, who used a radiation-hydrodynamics code to compute the time-dependent behavior of an axisymmetric disk structure and found that an outburst can produce circular surface waves of considerable amplitude that manifest themselves in scattered light images as bright and dark concentric rings.

The time-dependent nature of the heat propagation in nonequilibrium conditions may lead to peculiar features not seen in static radiation transfer. Among them are the apparent discrepancy between the central source luminosity and disk temperature; unusual vertical and radial temperature gradients; and  a time lag or full de-correlation between optical, infrared, and millimeter variability. As this behavior can be linked to observational phenomena such as temporal change in maser spots location \citep{2020NatAs...4..506B} or changing variability patterns with wavelength \citep{2017A&A...602A.120F, 2018ApJ...854...31J, 2020ApJ...897...54W,2020MNRAS.495.3614C}, it is important to study the effects of time-dependent radiation transfer (TDRT) to disentangle them from other potential explanations.

The disk chemical evolution during an outburst event is a much better explored topic both  theoretically and observationally. The major driver of chemical changes during the outburst is the sublimation of volatile species from dust grains, which leads to an increase in their gas-phase abundances by several orders of magnitude \citep{Banzatti_2012, 2015A&A...577A.102V, 2016ApJ...821...46T, 2017A&A...604A..15R, 2018ApJ...866...46M, 2019MNRAS.485.1843W,2024MNRAS.527.7652Z}. Some complex organic molecules were first detected in protoplanetary disks in outbursting sources, specifically \citep{2019NatAs...3..314L}, while molecule-rich spectra in currently quiet objects are supposed to be a signature feature of recently finished outburst activity \citep{2024MNRAS.527.7652Z,2024ApJ...975..170C}.

In the theoretical studies of outburst impact on the  physical and chemical structure of the disk, the temperature is often assumed to adapt instantaneously to the luminosity changes. In cases with low optical depths to the thermal radiation this approach is well justified \citep{2013ApJ...765..133J}, while for dense regions of protoplanetary disks the assumption may not hold. In the work by \citet{2024A&A...688A...8W}, closest to our paper, the TDRT was used to estimate the burst energetics without going into the details of the temperature distribution evolution. In addition, the outburst energetics and parameters of their object of interest (a massive young stellar object) differ greatly from those we consider in this paper. In the following sections, we study the time-dependent response of a protoplanetary disk to a $450\,L_{\sun}$ luminosity outburst event and focus both on the disk thermal evolution and potential observational manifestation.

\section{Physical model and methods}
\subsection{Time-dependent radiation transfer}

To calculate the thermal evolution of the disk we used the nonstationary FLD$^\text{s}$ method from~\cite{2024ARep...68.1045P} implemented in the numerical code HURAKAN~\citep{2022ARep...66..800P}. In this method the radiation field is split into stellar and dust thermal radiation. The heating by the stellar (central source) radiation is calculated using the frequency-dependent ray tracing method, while the diffusion approximation with a flux limiter (FLD approach) and the mean opacities are used to treat the propagation of the thermal radiation. Internal nonradiative heating such as viscous dissipation is also included via the source term. More details on the method, utilized opacities, and benchmarking are provided in Appendix~\ref{sec:AppA}.

The simulations are performed in 2D on a nonuniform polar~($R,\theta$) spatial grid with 200 radial and 200 angular cells. To trace density gradients, the radial grid is logarithmic and the polar grid is more condensed in the regions containing the bulk of the disk mass ($75\degr<\theta<105\degr$). The inner and outer radial grid boundaries are 0.03 and 250~au, the polar grid covers [0, $\pi$].

\subsection{Disk structure}
The radial disk structure is set by the gas surface density profile, which has two exponential cutoffs at the inner and outer characteristic radii, $R_{\rm in}$ and $R_{\rm out}$:
\begin{equation}
    \label{eq:Sigma}
    \Sigma_{\rm g} (R) = \Sigma_0 \left(\frac{R}{R_{\rm out}}\right)^{-\gamma} \exp \left[-\left(\frac{R}{R_{\rm out}}\right)^{2-\gamma} \right]  \exp \left[-\left(\frac{R}{R_{\rm in}}\right)^{\gamma-2} \right].
\end{equation}
The inner cutoff at $R_{\rm in} = 1$~au allows us to suppress disk self-shadowing by the inner wall. The values for the outer characteristic radius, disk, and stellar masses are set to roughly represent  the FU~Ori system itself ($R_{\rm out} = 11$~au, $M_{\rm disk}=6.6\times10^{-3}\ M_\odot$,  $M_{\star}=0.6\ M_\odot$; \citealt{2020ApJ...889...59P}). The assumed power law index $\gamma=1$ can be associated with a viscously evolving disk heated by central source radiation, resulting in a temperature profile $T\propto R^{-0.5}$ in the blackbody case.

Given the surface density, we calculate the volume density distribution in the meridional plane assuming the disk to be in hydrostatic equilibrium in the vertical direction. The temperature and vertical density profiles are iterated with each other to find pre-outburst stationary structure. After such initial temperature--density iterations, the density distribution is fixed for the outburst and post-outburst thermal relaxation. While the gas density is likely affected by the luminosity outburst as well, in this paper we focus purely on disk thermal evolution. Understanding the time-dependent temperature evolution on a fixed density structure is crucial for future studies of the coupled thermal and hydrodynamical disk response.

In our model, the central source luminosity $L_{\nu}$ is the sum of two blackbody components representing the intrinsic stellar and accretion-induced radiation:
\begin{equation}
    L_{\nu} (t) = 4\pi R_{\star}^2 \cdot \pi B_{\nu}(T_{\star}) +\frac{\pi B_{\nu}(T_{\rm acc})}{\sigma T_{\rm acc}^4}  L_\text{acc}^\text{UV}(t).
    \label{Lnu_t}
\end{equation}
The stellar component is not evolving ($R_\star = 3.6\, R_\odot$, $T_\star = 4000$~K, $L_{\star}=3\,L_{\sun}$; \citealt{2015A&A...577A..42B}), while the accretion source with effective temperature $T_{\rm acc} = 10^4$~K is variable according to a prescribed outburst luminosity curve $L_\text{acc}^\text{UV}(t)$.

The physical mechanism(s) behind the FU Ori outburst events is a matter of debate. To cover a range of possibilities we consider two models depending on where the sudden accretion rate increase is confined: within the innermost active disk on sub-au scales with the outer disk beyond 1 au being stable or when the whole radial disk extent is affected by accretion rate jump. In the first model, which we refer to as the passive outer disk model or Model P, the whole outburst energy, which we set to $450\,L_{\sun}$ at maximum, is released in the star vicinity and affects the outer disk only via the UV/optical radiation. In the second fully active disk model (Model A), we assume that the whole disk experiences an instantaneous increase in accretion rate $\dot{M}$ from a background value of $2\times10^{-8}\ M_{\sun}\,\text{yr}^{-1}$ to a much higher rate, which is constant over the disk. The viscous heat is fully neglected in Model P to make the comparison between the models  clearer.

Most of the heating of the outer disk in Model A is still caused by the strong UV/optical radiation from the inner regions of the active disk inside some radius $R_\text{A}$. The outer regions of the active disk (beyond $R_\text{A}$) are heated both viscously and radiatively. Separating the active disk into two parts allows us to elaborate a procedure keeping the total outburst energetics the same between the models allowing their direct comparison (see Appendix \ref{sec:AppB} for details). For our assumed value of $R_\text{A}=0.1$~au, the total outburst energetics at maximum in  Model A  ($450\,L_{\sun}$) is split into the radiative part of $412\,L_{\sun}$ and the viscous part of $38\,L_{\sun}$ released directly in the disk.

\section{Results}
\begin{figure*}[!ht]
    \centering
    \begin{minipage}[!ht]{\textwidth}
         \center{Passive outer disk (Model P) \includegraphics[width=1\linewidth]{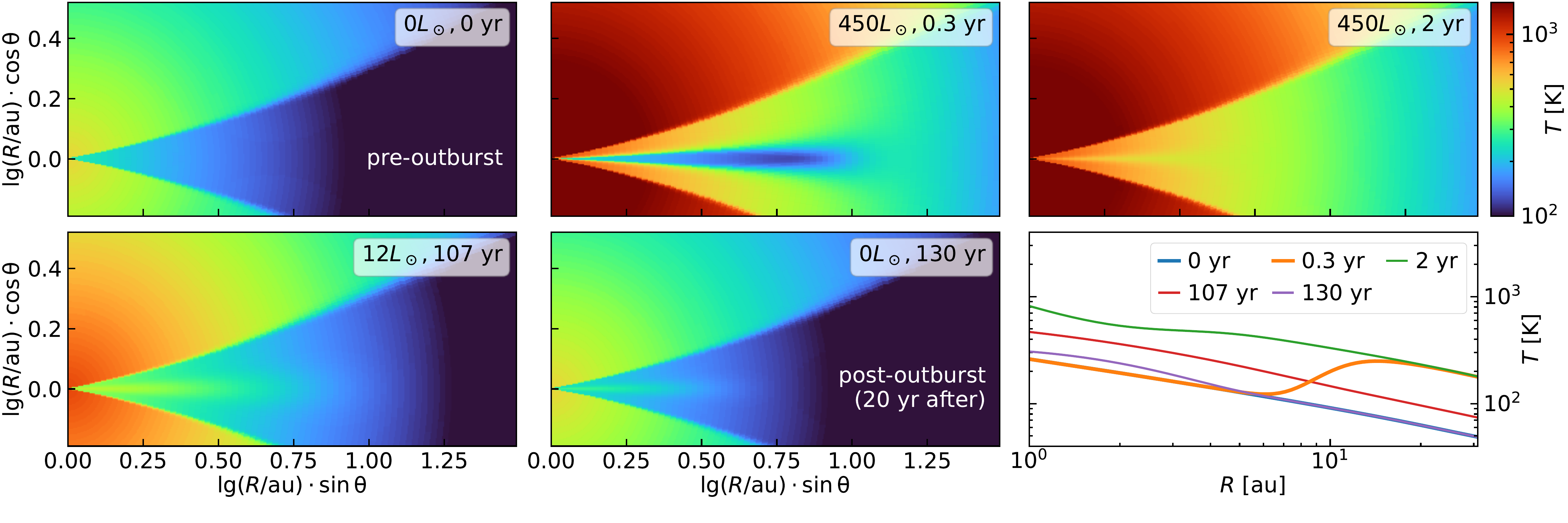}} \\ 
    \end{minipage}
    \vspace{3pt}
    \vfill
    \begin{minipage}[!ht]{\textwidth}
        \center{Fully active disk (Model A) \includegraphics[width=1\linewidth]{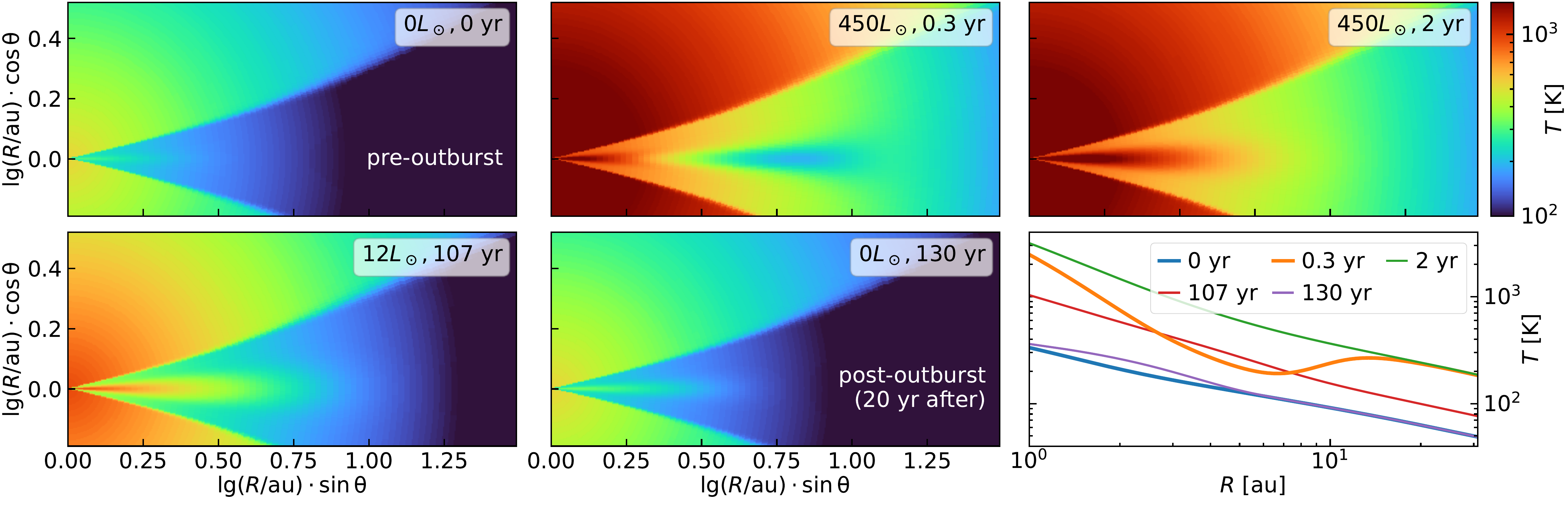}} \\
    \end{minipage}
    \caption{Evolution of the disk thermal structure during the outburst for the passive outer disk model (Model P; upper set of panels) and the fully active disk model (Model A; lower set of panels). Each panel set shows a selection of time moments; the total outburst energetics are shown in the legend of each panel. The lower right panel in each set presents the midplane temperature radial profile for the same time moments. The chosen coordinates for the heatmaps $\lg (R/\mbox{au}) \sin \theta$ -- $\lg (R/\mbox{au}) \cos \theta$ allow us to highlight the inner and midplane regions and are convenient for representing data obtained on the polar $(R, \theta)$ grid.}
    \label{pic:snapshots}
\end{figure*}

The energy radiated during the FU Ori type outburst greatly exceeds the total disk thermal energy at the pre-outburst stage. It still takes time for the radiative flux to diffuse toward the dense disk midplane and be converted to gas thermal energy. Local energy dissipation such as viscous heating takes time as well because of nonzero gas heat capacity and the finite timescales required to equilibrate the thermal radiation field in optically thick conditions. As a consequence, one may expect nonhomogeneous disk temperature evolution during the outburst in both the radial and vertical directions. To characterize this evolution we consider time-dependent radiation transfer using the HURAKAN code for two models described above: the passive outer disk (Model P) and the fully active disk (Model A).

At the pre-outburst state, we run the simulations for $10^4$~yr to get a quasi-stationary structure corresponding to the stellar luminosity $L_{\star}=3\,L_{\sun}$. At $t = 0$ the radiative accretion source $L_\text{acc}^\text{UV} = 450\,L_{\sun}$ is initiated for Model P. For Model A, the corresponding value of $L_\text{acc}^\text{UV}$ is smaller ($412\,L_{\sun}$; see Equation~\ref{eq:B3}) as the rest is deposited via viscous heating with the total power of $L_\text{acc}^\text{disk} = 38\,L_{\sun}$ corresponding to $\dot{M}= 9\times10^{-5} M_{\sun}\,\text{yr}^{-1}$ (Equation~\ref{eq:gamma_heat}). These values are linearly scaled from their maximum values to zero in 110 years. Although the actual luminosity decline may deviate from a linear temporal dependence and vary between objects, we adopt this basic case for simplicity. The profiles for $L_{\rm acc}^{\rm UV}$ and their comparison with the observational data for FU Ori are shown in Fig.~\ref{pic:Lacc}.

The 2D temperature distributions for a set of time moments catching pre-outburst ($t=0$), early (0.3 and 2~yr), late (107~yr), and post-outburst (130~yr) stages are shown in Fig.~\ref{pic:snapshots}. For  Model P, shown in the upper set of snapshots, one can see that the sudden rise in the central source luminosity leads to immediate heating of the disk atmosphere, while the disk midplane inside 10~au is almost unaffected. At the next time moment of $t=2$~yr (right panel in the first row of Fig.~\ref{pic:snapshots}), when the central source luminosity is still at its highest value, the cool midplane zone shrinks in both radial and vertical extents. This zone is fully heated only after $t=5$~yr, when the accretion luminosity already starts to drop (not shown in these panels), which corresponds to the characteristic timescales shown in Fig.~\ref{pic:FigC}. As the accretion luminosity declines during later outburst stages (left panel, second row of Fig.~\ref{pic:snapshots}, $L_{\rm acc}^\text{UV}=12\,L_{\sun}, t=107$~yr), the same physical mechanism,   where optically thin regions adapt more quickly to luminosity changes,  produces the inverse temperature distribution. The disk midplane keeps the heat for a decade after the end of the outburst, and is thus hotter than the upper and outer disk parts.  The right panel of the second row summarizes the evolution of the temperature profile in radial direction in the midplane.

In the fully active disk (Model A), the viscous dissipation produces an overheated midplane region inside several au. At 0.3 yr, the radial temperature profile exhibits a minimum at a location similar to that in Model P. Later, at 2~yr the viscously heated zone spreads outward and engulfs the underheated region. This peculiarity in the thermal structure may have consequences for the disk chemistry during early months of outburst monitoring. Later and post-outburst stages of  Model A are qualitatively similar to those of Model P with the disk midplane slowly releasing the heat accumulated during the intense heating phase.

\begin{figure*}[!ht]
   \centering
   \includegraphics[width=0.44\textwidth]{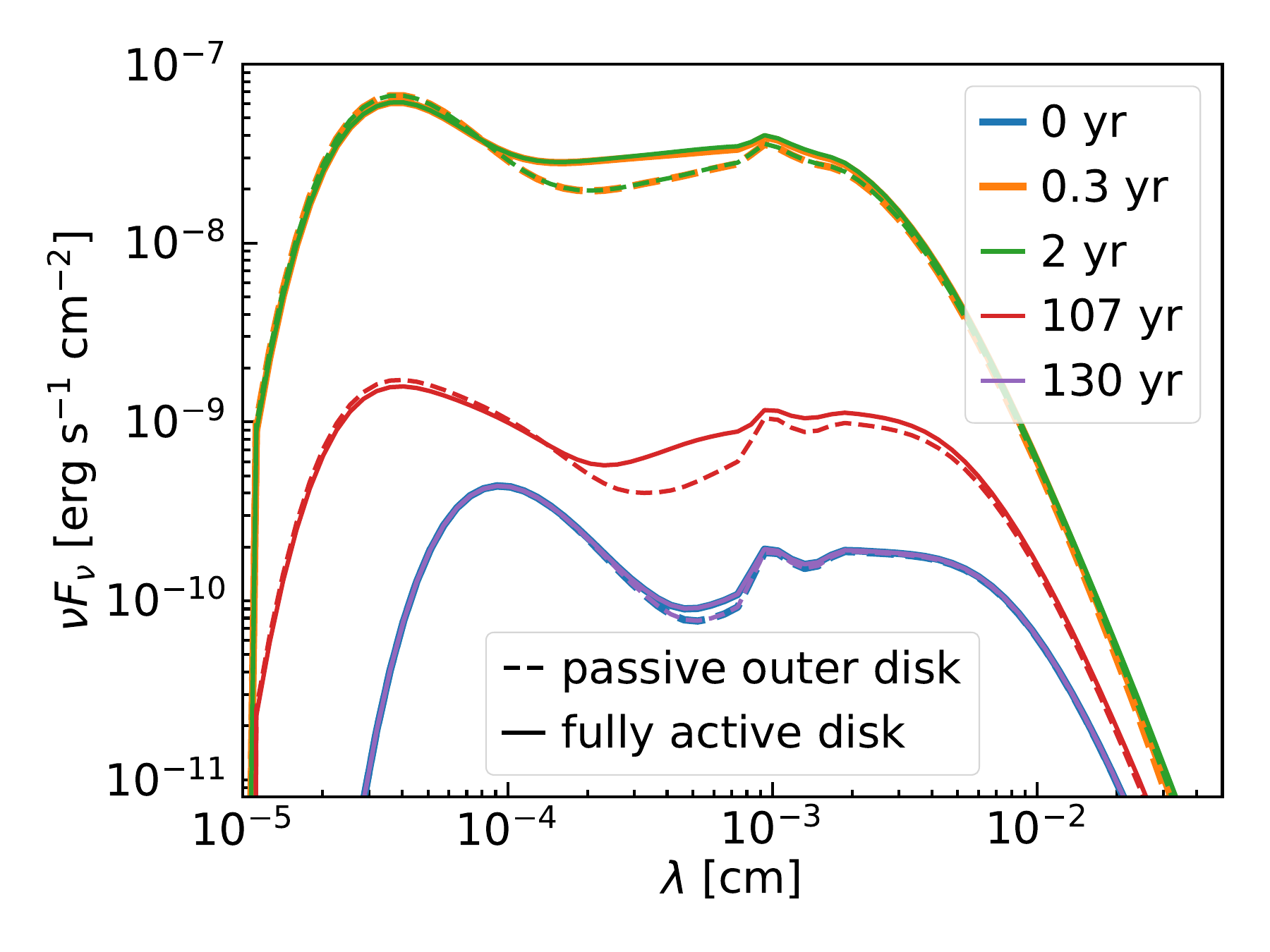}
   \includegraphics[width=0.48\textwidth]{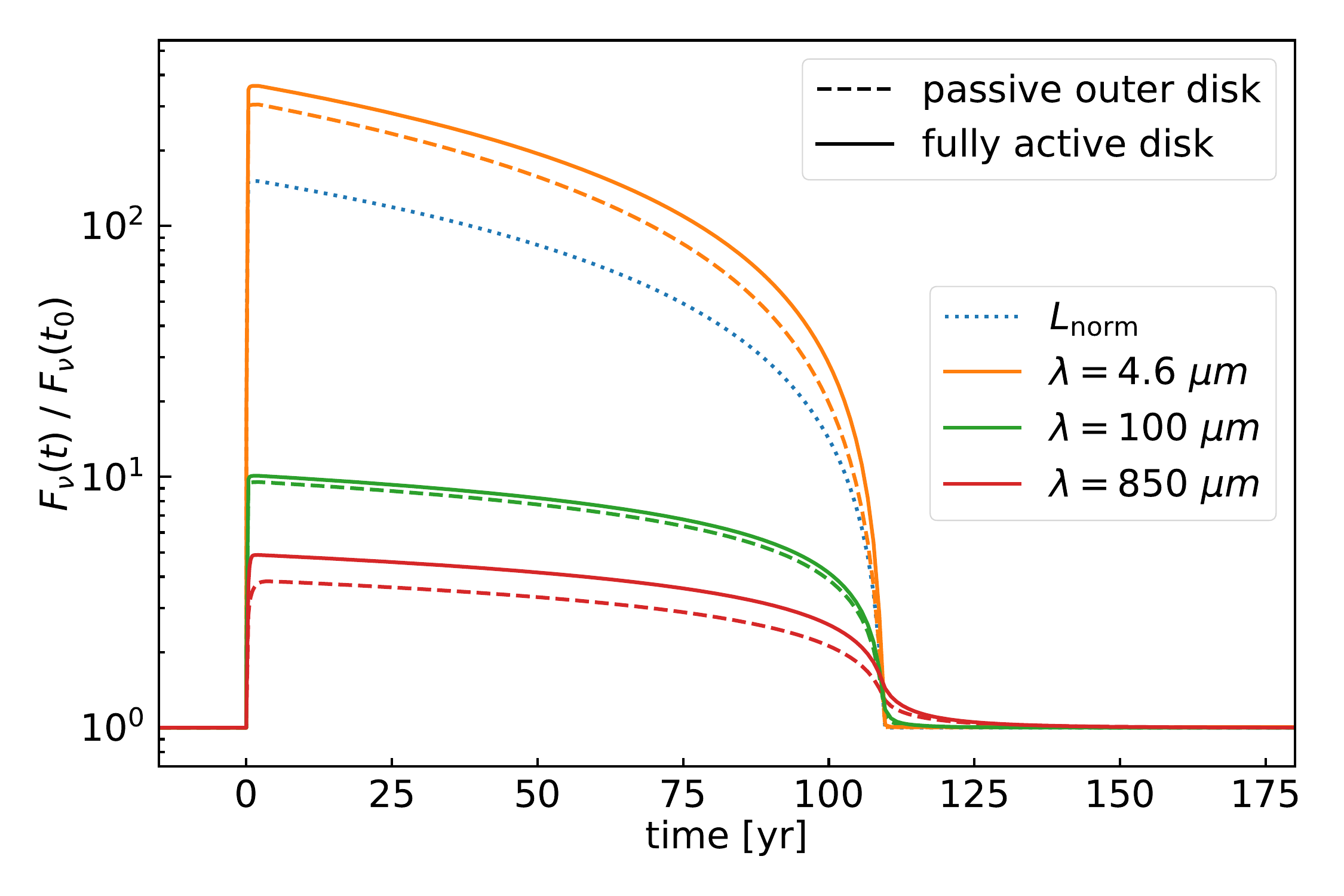}
\caption{Left panel: Spectral energy distributions for different stages of the outburst. The profiles for Model~P and Model~A are shown with dashed and solid lines, respectively. Right panel: Evolution of the corresponding radiation fluxes for several wavelengths. The reference ratio $(L_\star + L_\text{acc}^\text{UV}) /L_\star$ for Model P is shown with a dotted line.}
\label{pic:Fig2}
\end{figure*}

To study possible observable effects, we calculate the corresponding SED evolution using RADMC-3D code \citep{2012ascl.soft02015D}, adopting a distance of $d=402$~pc \citep{2023A&A...674A...1G} and a disk inclination of $i=37\degr$ \citep{2020ApJ...889...59P}. As the current version of RADMC-3D does not include time-dependent radiation transfer, we use it only for SED calculations based on the temperature distribution obtained in HURAKAN simulations. As shown in Fig.~\ref{pic:Fig2} the SED varies significantly during the outburst. The SED maximum related to the central source radiation moves toward the shorter wavelengths (from 1 to $0.3\,\mu$m) during the increase in luminosity. The SED profile between $1$ and $10\,\mu$m becomes flatter at high luminosities, so the silicate feature at 10 $\mu$m becomes less pronounced. The differences between the outer passive disk model and the fully active disk model are also visible. The presence of warmer dust in the midplane of the fully active disk model results in higher fluxes at $1-10\,\mu$m. In turn, the flux at shorter wavelengths is reduced in Model A due to the lower luminosity of central source. The effect of nonstationary radiative transfer is not very pronounced in the SEDs. The differences between SEDs for time moments $t=0.3$~yr and $t=2$~yr with the same accretion luminosity of 450~$L_\star$ and for time moments $t=0$~yr and $t=130$~yr (before the outburst and 20~years after it) are negligible.

In the right panel of Fig.~\ref{pic:Fig2} we also show the evolution of the normalized total fluxes $F_{\nu}(t)/F_{\nu}(0)$ calculated for three wavelengths $\lambda = 4.6,100,850\ \mu \rm m$. As one can see, the fluxes follow the shape of the normalized accretion luminosity curve, but with different scale factors. The flux increase at $4.6\,\mu$m is higher than the increase in total luminosity. In turn, the flux increase at $\lambda = 100$ and $850\ \mu \rm m$ is much lower than the total luminosity raise. It should be noted that only a fraction of the radiative heat is  intercepted by the disk (typically, 35-40\%), but all the dissipative heat goes to the disk. While the total (radiative plus dissipative) outburst energetics is the same for the two models, the total energy deposited to the disk is smaller in Model P than in Model A because of a higher amount of radiation in  Model P 
freely escaping the system through polar regions. This explains why the IR fluxes for Model P are slightly lower than for Model A, as shown in the right panel of Fig.~\ref{pic:Fig2}.

The effects of nonstationary radiation transfer is mostly hidden in the dust continuum observations as the regions with time-dependent temperature anomalies are in the optically thick regions to disk thermal radiation. The nonstationary effects are only noticeable at (sub-)millimeter wavelengths for a decade after the end of the outburst (between 110 and 125~yr). 

\section{Conclusions}

Using the time-dependent radiative transfer, we considered the impact of an FU~Ori type luminosity outburst on the thermal structure of the protoplanetary disk. To cover possible physical mechanisms driving the outburst, we examined two scenarios: one in which the increased accretion rate is confined to the compact sub-au inner region with the outer disk being fully passive (Model P) and one where the outer disk is fully active  (Model A).

We find that the protoplanetary disk heats up and cools down nonuniformly in the radial and vertical directions. Optically thick disk regions in both scenarios require several years to become fully heated during the outburst and a decade to cool after it. The upper layers and outer parts of the disk, which are optically thin to thermal radiation, are heated and cooled almost instantaneously. This creates an unusual radial temperature profile during the early heating phase with a minima at several au for both scenarios. During the cooling phase, an inverse temperature gradient in the vertical direction -- with upper layers colder than the midplane -- appears even in the passive disk model. The fully active Model A is characterized by a strong dissipative energy source inside several au, which speeds up the midplane heating at early outburst stages.

Models P and A reveal noticeable differences in their SEDs at $1-10\,\mu$m due to the presence of warmer dust in Model A. As the nonstationary effects are confined to the disk regions that are optically thick to the thermal radiation, these effects are mostly hidden from the dust continuum observations.

The inhomogeneously evolving disk temperature is likely to affect other key aspects of disk physics, including the morphology of snowlines and the ionization structure. Consequently, further investigation of disk chemical evolution under nonequilibrium conditions -- along with the corresponding molecular emission signatures -- represents a particularly interesting area for future research.

\begin{acknowledgements}
    The authors are grateful to the reviewer for constructive suggestions that helped us to improve the content of the paper.
    This work was also supported by the NKFIH NKKP grant ADVANCED 149943 and the NKFIH Excellence Grant TKP2021-NKTA-64. Project no.149943 has been implemented with the support provided by the Ministry of Culture and Innovation of Hungary from the National Research, Development and Innovation Fund, financed under the NKKP ADVANCED funding scheme.
\end{acknowledgements}

\bibliographystyle{aa}
\bibliography{bibfile}

\appendix
\section{Details of protoplanetary disk thermal model}
\label{sec:AppA}
To calculate the thermal evolution of a protoplanetary disk we adopt the FLD$^\text{s}$ method from~\cite{2022ARep...66..800P}. Here we summarize the model. The governing equations are
\begin{eqnarray}
\rho c_{\rm V} \frac{\partial T}{\partial t}&=& c \alpha_{\rm P} \left(E - a T^4\right) +S^\text{UV} + S_\text{acc}^{\rm disk},
\label{therm_sys1}\\
\frac{\partial E}{\partial t}&=& - c \alpha_{\rm P} \left(E - a T^4\right) + \hat{\Lambda} E,
\label{therm_sys2}
\end{eqnarray}
where $\rho$ is the gas volume density, $c_{\rm V}$ is its specific heat capacity [erg g$^{-1}$ K$^{-1}$], $c$ is the speed of light,
$\alpha_{\rm P}$ [cm$^{-1}$] is the Planck-average opacity coefficient, $S^\text{UV} = S_{\star} + S_\text{acc}^\text{UV}$ [erg cm$^{-3}$ s$^{-1}$] is the heating rate by central source radiation consisting of the stellar and accretion components, and $S_\text{acc}^\text{disk}$ [erg cm$^{-3}$ s$^{-1}$] is the heating rate due to internal disk dissipation, $T$ is the temperature of the medium (equal for gas and dust), and $E$ is the energy density of thermal radiation.

Equation \eqref{therm_sys1} describes the change in the thermal energy of the medium due to the absorption and re-emission of thermal radiation (terms $c\alpha_{\rm P}E$ and $c\alpha_{\rm P}a T^4$, respectively), due to the absorption of direct central source radiation ($S_{\star}$), and due to internal accretion heating ($S_\text{acc}^\text{disk}$). Equation \eqref{therm_sys2} describes the change in the energy density of radiation due to the absorption and re-emission of thermal radiation, and due to the spatial diffusion of
thermal radiation, represented by the operator $\hat{\Lambda}E$,
\begin{equation}
\hat{\Lambda}E = -{\nabla}\cdot {\bf F}
= {\nabla} \left( \frac{c\lambda}{\alpha_{\rm R}}{\nabla}\, E \right),
\label{eq:operE}
\end{equation}
where ${\bf F}$ is the flux of thermal radiation, $\alpha_{\rm R}$ is the
Rosseland-average opacity coefficient, and
$\lambda$ is the flux limiter. The calculation of $\lambda$ is
performed according to the flux limited diffusion (FLD)
theory \citep{1981ApJ...248..321L}.

\begin{figure}[!ht]
   \centering
   \includegraphics[width=0.45\textwidth]{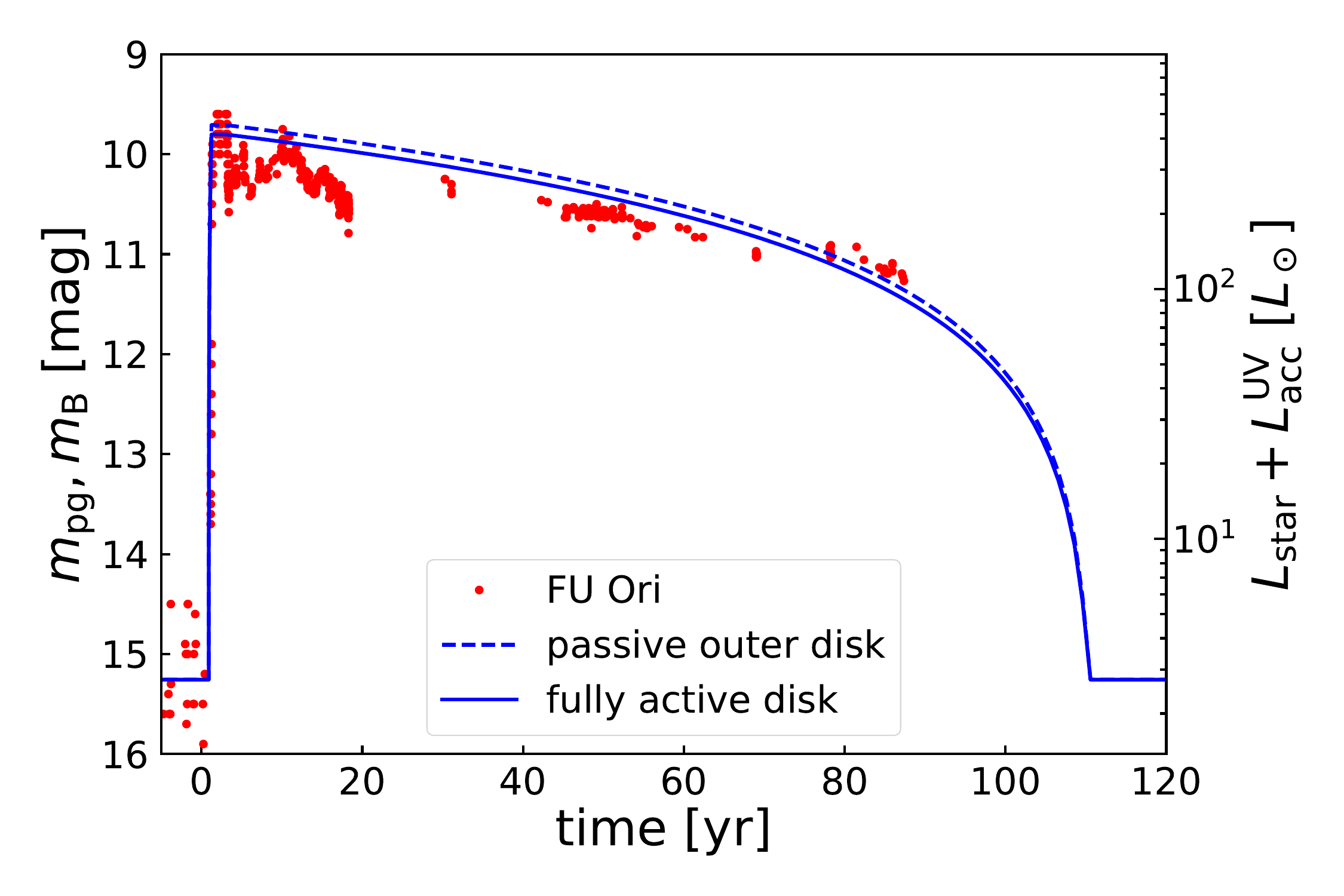}
   \caption{Central source luminosity profiles $L_{\star} + L_{\rm acc}^\text{UV}$ used in time-dependent radiation transfer for the outer passive disk and fully active disk models  (blue lines; the vertical axis on the right). 
   The observed light curve of FU Ori, incorporating data from  \citet{1939BHarO.911...41H, 1954ZA.....35...74W, 1968ApJ...151..977M, 1985SvAL...11..358K, 1988ApJ...325..231K, Kenyon_2000, Green_2006, 2018A&A...618A..79S, 2018MNRAS.477.3145J}, and additional data for 2017--2023 ($t>80$~yr) obtained from the Schmidt and RC80 telescopes at Konkoly Observatory is shown by dots.}
   \label{pic:Lacc}
\end{figure}

The heating function due to central source radiation $S^\text{UV}$
is calculated as
\begin{equation}
S^\text{UV}=-\nabla \cdot {\bf F}^\text{UV},
\label{eq:stellarheat1}
\end{equation}
where {${\bf F^\text{UV}}$} is the central source radiation flux integrated over frequency:
\begin{equation}
{\bf F}^\text{UV} = \int\limits_0^{\infty}  {\bf F}^\text{UV}_{\nu}\, d\nu.
\label{eq:stellarheat2}
\end{equation}
The spectral flux ${\bf F}^\text{UV}_{\nu}$ is calculated taking into
account the true absorption of radiation from the star and active inner disk to the considered
medium element along the radial direction. In this case the radial
component of the flux is
\begin{equation}
F^\text{UV}_{\nu} =  \frac{L_{\nu} e^{-\tau_{\nu}(R)}}{4\pi R^2},
\label{eq:stellarheat3}
\end{equation}
where $L_{\nu} = L_{\nu}(t)$ is defined by Equation~(\ref{Lnu_t}), and $\tau_{\nu}(R)$ is the optical depth at frequency $\nu$ along the line of sight from the star to the considered
point. For simplicity, we neglect scattering for calculation of ${\bf F}^\text{UV}_{\nu}$. For the outburst models presented in this paper we assume the temporal behavior for the total luminosity $\int L_{\nu}(t) d\nu = L_\star + L_\text{acc}^\text{UV}$, as shown in Fig.~\ref{pic:Lacc}.

The system of equations~\eqref{therm_sys1}--\eqref{therm_sys2} must be
supplemented with boundary conditions.  Thermal radiation from the inner boundary of the disk may partially escape via the polar regions. The amount of radiation
returning to the disk from opposite walls depends on the height,
position of the inner wall, and other disk parameters, and
is difficult to predict in general. At the inner boundary of the spatial grid, we
use the condition that the radiation flux is proportional to the product
of energy density and the speed of light,
\begin{equation}
\left. F\right|_{\text{in}} = -pc \left(E-E_\text{cmb}\right),
\label{eq_bc2}
\end{equation}
where $F$ is the radial component of the flux determined by
Equation~\eqref{eq:operE}, and $E_{\text{cmb}}$ is the energy density
of the cosmic microwave background. The coefficient $p$ is introduced phenomenologically
and describes the fraction of freely escaping radiation. The value $p=1/2$ corresponds
to the case where radiation isotropically escapes the medium. The coefficient $p=1$
(used in our modeling) corresponds to the limiting case where
radiation freely escapes the medium perpendicular to the boundary.

The outer boundary conditions are set assuming free escape of the thermal radiation from the computational domain:
\begin{equation}
\left. F\right|_{\text{out}} = c \left(E-E_\text{cmb}\right).
\label{eq_bc3}
\end{equation}

The numerical method to solve Equations~\eqref{therm_sys1}--\eqref{therm_sys2}
is based on an implicit finite-difference scheme constructed in spherical
coordinate system. The linearization of finite-difference equations allows
to reduce the problem to a system of linear algebraic equations with a sparse
matrix. To invert the matrix a modified Gauss method is adopted. More
details on numerical method can be found in~\cite{2022ARep...66..800P}.

In our model, it is assumed that the only source of opacity is dust, and
the temperatures of the gas and dust are equal. The ratio of dust density
to gas density throughout the disk is assumed to be constant and equal to
0.01, i.e., the dust is considered uniformly mixed with the gas. The
coefficients $\alpha_{\rm P}=\rho\kappa_{\rm P}$ and $\alpha_{\rm
R}=\rho\kappa_{\rm R}$ appearing in the equations above, are determined
via the Planck and Rosseland mean opacities, 
$\kappa_{\rm P}$ and $\kappa_{\rm R}$ [cm$^2$ g (gas)$^{-1}$].
The required Planck and Rosseland mean dust opacities are calculated using frequency-dependent absorption coefficients for a mixture of spherical silicate and carbonaceous dust particles (the mass fraction of carbonaceous dust particles is 0.2; the size distribution is $n\propto a^{-3.5}$ with minimal and maximal radii of the dust particles being $0.005$ and $1\,\mu$m, respectively). The frequency-dependent absorption coefficients themselves are calculated using the Mie theory~\citep{2004CoPhC.162..113W}. The mean and frequency-dependent opacities 
for the considered dust grain ensemble are shown in Fig.~1 of \cite{2024ARep...68.1045P}.
We take 5/3 for the value of the gas heat capacity ratio, which represents molecular hydrogen at low temperatures and 3:1 ortho-para ratio ($\lesssim 150$~K) as well as helium. The assumed mean molecular weight is 2.3.

\begin{figure}[!ht]
  \centering
  \includegraphics[width=0.45\textwidth]{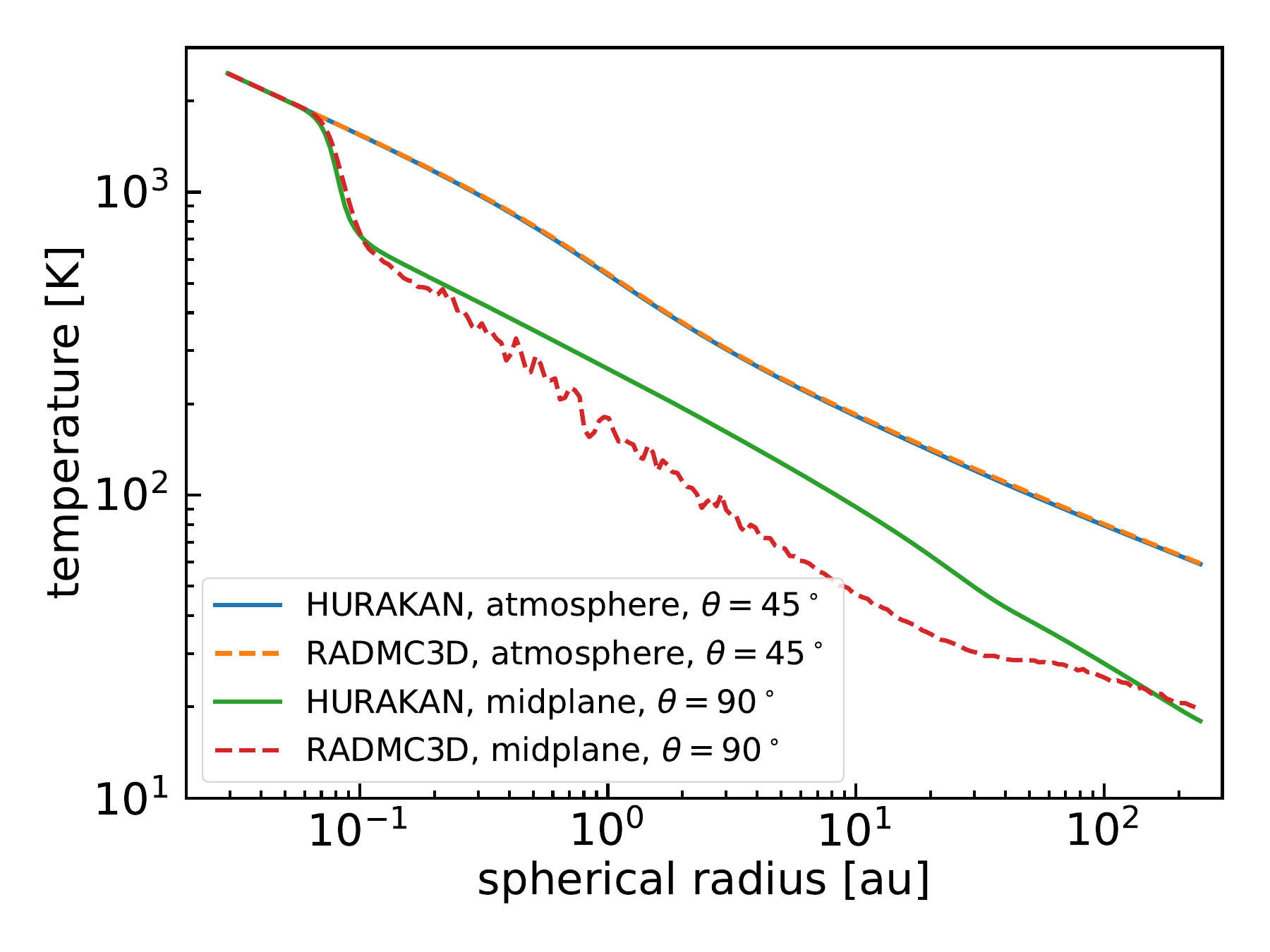}
  \caption{Comparison between temperatures calculated using HURAKAN (FLD$^\text{s}$) and RADMC-3D (Monte Carlo) in the disk atmosphere (polar angle $\theta=45\degr$) and midplane ($\theta=90\degr$)  for stationary pre-outburst stage.}
  \label{pic:bench}
\end{figure}

Owing to the consideration of UV/optical heating via the $S^\text{UV}$ term, the FLD$^\text{s}$ method implemented in HURAKAN provides accurate calculation of dust temperature in the upper disk layers that are optically thin for thermal radiation. The use of Planck and Rosseland opacities speeds up the calculation of the time-dependent radiation transfer and allows for its use in hydrodynamical simulations. However, the spectrally averaged opacities and the diffusion nature of the FLD approach lead to a loss of accuracy of temperature calculation in anisotropic setups like protoplanetary disks. In Fig.~\ref{pic:bench} we show the radial profiles of disk temperatures calculated with HURAKAN and RADMC-3D at pre-outburst stage for Model P. As HURAKAN radiation transfer module is time-dependent, we assume a long time-step of $10^4$~yr, which is larger than thermal timescales, to match the stationary approach of RADMC-3D. While the temperature profiles in the disk upper layers (the polar angle $\theta=45\degr$) calculated with the two codes almost ideally correspond to each other, the FLD temperatures are notably higher in the midplane. The difference is caused by both the spectral averaging of opacities and anisotropic disk conditions \citep{2002A&A...389..464D, 2025arXiv250410220P, 2025arXiv250413999M}. More details on the numerical tests and benchmarking between HURAKAN and RADMC-3D codes in stationary approximation are presented in~\cite{2024ARep...68.1045P} and~\cite{2025arXiv250410220P}. 

\section{Calculating the heating in fully active disk model}
\label{sec:AppB}
 We assume that luminosity during the accretion event is provided mostly by the release rate of gravitational energy $L_\text{grav}=GM_{\star}\dot{M}/R_{\star}$, so the accretion rate $\dot{M}$ onto the star is:
\begin{equation}
\dot{M}=\frac{L_\text{grav}R_{\star}}{GM_{\star}},
\end{equation}
where $M_{\star}$ and $R_{\star}$ are the stellar mass and radius. For the fully active disk model we assume that the accretion rate is constant over the disk, equals to that onto the star and results in a time-independent gas surface density (contrary to the passive outer disk model where the accretion energy is assumed to be released in a compact inner region providing only UV/optical heating to the outer passive disk).

We assume that the heating rate of the disk beyond some radius $R_\text{A}$ is equal
to the release rate of gravitational energy,
\begin{equation}
L_\text{acc}^\text{disk} = \frac{1}{2}\frac{GM_{\star}\dot{M}}{R_\text{A}},
\label{eq:Ldisk}
\end{equation}
where the coefficient ($1/2$) accounts for the fact that half of the gravitational energy is converted into Keplerian motion of the gas.

In our approach, the disk is split into two parts at some radius $R_\text{A}$. The inner part of the active disk, $R<R_\text{A}$, is so hot that its UV/optical radiation is treated as the contribution to the central luminosity source. In the outer part of active disk $R>R_\text{A}$ the accretion heating contributes mostly to the thermal radiation and has the energy release rate of $L_\text{acc}^\text{disk}$. The sum of the energy release rates in two parts of active disk provides the gravitational energy release rate $L_\text{acc}^\text{UV} + L_\text{acc}^\text{disk} = L_\text{grav}$.
The accretion contribution to central source luminosity can be rewritten as
\begin{equation}\label{eq:B3}
L_\text{acc}^\text{UV} = L_\text{grav} \left( 1-\frac{R_\star}{2R_\text{A}} \right).
\end{equation}
The heating function $\Gamma_\text{acc}^\text{disk}$ per unit disk area which
is consistent with Equation~\eqref{eq:Ldisk}, i.e.,
$L_\text{acc}^\text{disk} = \int\limits_{R_\text{A}}^{\infty} \Gamma_\text{acc}^\text{disk}\, 2\pi R\, dR$,
has the form
\begin{equation}
\Gamma_\text{acc}^\text{disk} = \frac{1}{4\pi} \frac{GM_{\star}\dot M}{R^3}.
\label{eq:gamma_heat}
\end{equation}
The heating rate per unit volume due to internal accretion heating is finally calculated as
\begin{equation}
S_\text{acc}^{\rm disk}=\rho\frac{\Gamma_\text{acc}^\text{disk}}{\Sigma_\text{g}},
\end{equation}
where $\Sigma_\text{g}$ is the surface density of the disk.

In our prescription of active disk model we do not use the standard formalism of the viscous disks, which was initially developed for black holes. In the stationary case, the heating rate due to viscous dissipation is usually written as~\citep{1981ARA&A..19..137P}
\begin{equation}
\Gamma_\text{vis}=\dfrac{3}{4\pi}\dfrac{GM_{\star}\dot M}{R^3} \left[1-\left(\dfrac{R_\star}{R}\right)^{1/2}\right],
\label{eq:gamma_vis}
\end{equation}
which vanished in the stellar vicinity. This rate is coupled with the equation for the stationary surface density distribution,
\begin{equation}
\nu\Sigma_\text{g} = \dfrac{\dot{M}}{3\pi} \left[1-\left(\dfrac{R_\star}{R}\right)^{1/2}\right] ,
\label{eq:sigma}
\end{equation}
where $\nu$ is the kinematic viscosity. Combining Equations \ref{eq:gamma_vis} and \ref{eq:sigma} provides the classical heating rate $\Gamma_\text{vis}=(9/4)\nu\Sigma_\text{g}\Omega_{\rm K}^2$ \citep{1981ARA&A..19..137P}.

The problem of using Equations~\eqref{eq:gamma_vis} -- \eqref{eq:sigma} is that they are derived assuming a zero-torque boundary condition, which can be less appropriate for protoplanetary disks than for black holes. In particular, Equation \eqref{eq:sigma} forces the surface density to vanish at the stellar boundary $R_\star$ resulting in an infinite radial velocity at this location to keep the fixed value of $\dot M$. The outcome of the zero-torque boundary is also that viscous heating rate \eqref{eq:gamma_vis} is zero at $R_\star$ but is three times higher than the rate of gravitational energy release \eqref{eq:gamma_heat} at $R\gg R_\star$.  Given these circumstances,  we prefer to use  Equation \eqref{eq:gamma_heat} to satisfy local energy conservation relations but neglecting possible nonlocal effects of turbulent viscosity.

\section{Characteristic timescales for disk midplane radiative heating and cooling}

The energy balance equation can be written per unit surface as
\begin{equation}
  \Sigma_{\rm g}\frac{d u}{d t}=\Sigma_{\rm g}c_V\frac{d T}{d t}=\Gamma - \Lambda,
\label{eq:Sigmadudt}
\end{equation}
where $u$ [erg~g$^{-1}$] is the specific internal energy of the gas, related to the specific heat capacity $c_V$ as $u = c_V T$ for temperature independent $c_V$, $\Gamma$ and $\Lambda$ are  the heating and cooling functions, respectively. According to~\cite{Pavlyuchenkov_2023ARep...67..470P}, they can be approximated as
\begin{equation}
 \Lambda = \dfrac{4 \tau_{\rm P} \sigma T^4}{1 + 2\tau_{\rm P} + \dfrac{3}{2}\tau_{\rm P}\tau_{\rm R}}, \ \ 
 \Gamma = \dfrac{2 \mu_0 F_0 \tau_{\rm P}}{1 + 2\tau_{\rm P} + \dfrac{3}{2}\tau_{\rm P}\tau_{\rm R}},
\label{eq:LG}
\end{equation}
where $T$ is the midplane temperature, $\tau_{\rm P}$ and $\tau_{\rm R}$ are  the Planck and Rosseland  optical depths to the disk midplane equal to $(1/2)\kappa_\text{P} \Sigma_\text{g}$ and $(1/2)\kappa_\text{R}\Sigma_\text{g}$, respectively, $\mu_0$ is the cosine of the angle between the direction to the star and the normal to the disk surface (taken as 0.05), and $F_0$ is the stellar radiative flux incident on the disk surface at a given radius.

To estimate the characteristic heating  timescale  $t_{\rm heat}$ of the disk midplane, let us consider a sharp increase in disk heating, i.e., $\Lambda \ll \Gamma$. By substituting Equation~\eqref{eq:LG} into \eqref{eq:Sigmadudt} and solving for time (for estimates, we use $\dfrac{d T}{d t} \approx \dfrac{\Delta T}{\Delta t}$ and $\Delta T \sim T$), we derive the disk heating time $t_{\rm heat}$ as a function of the radial position in the disk,
\begin{equation}
 t_{\rm heat} = \frac{c_V T_0 (1 + 2 \tau_{\rm P} + 1.5\tau_{\rm P} \tau_{\rm R})}{2 \mu_0 \kappa_{\rm P} F_0},
\label{eq:t_h}
\end{equation}
where $T_0$ is the midplane temperature of the disk at the time just before the outburst.

Similarly, in the case of a sharp decline of disk heating ($\Lambda \gg \Gamma$), the disk cooling timescale $t_{\rm cool}$ is derived as
\begin{equation}
 t_{\rm cool} = \frac{c_V (1 + 2 \tau_{\rm P} + 1.5\tau_{\rm P} \tau_{\rm R})}{4 \kappa_{\rm P}  \sigma T_{\rm max}^3},
\label{eq:t_c}
\end{equation}
where $T_{\rm max}$ is the maximum midplane temperature attained during the outburst, for which we take the moment of $t=2$~yr.

\begin{figure}[!ht]
\centering
 \includegraphics[width=0.45\textwidth]{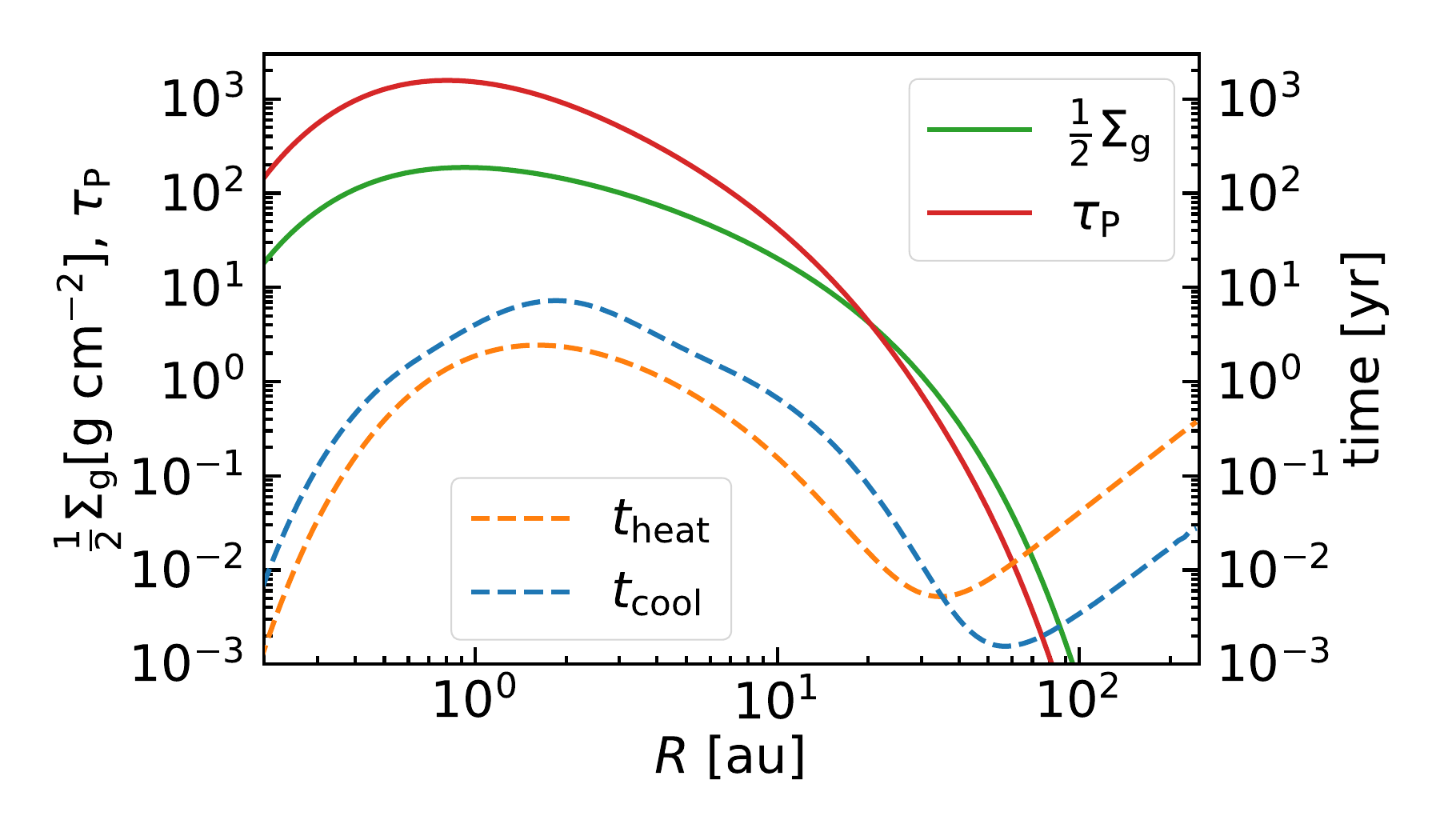}
 \caption{Characteristic timescales for the disk midplane heating (at $t=0$) and cooling (at $t=2$~yr) for Model P shown with the dashed lines. The gas surface density and the Planck optical depth to the disk midplane are shown with the solid lines.}
\label{pic:FigC}
\end{figure}

The resulting characteristic timescales for Model P as a function of disk radial position are shown in Fig.~\ref{pic:FigC}, the radial profiles of the gas surface density $\Sigma_{\rm g}/2$ to the disk midplane and the corresponding Planck optical depth $\tau_{\rm P}$  are also plotted. These analytical estimates are consistent with the full simulations shown in Fig.~\ref{pic:snapshots}.

\end{document}